\documentclass[runningheads]{llncs}
\usepackage[scaled]{beramono}
\usepackage[T1]{fontenc}
\usepackage{graphicx}
\usepackage{hyperref}

\usepackage[dvipsnames]{xcolor}

\usepackage{alloy}
\usepackage{adjustbox}

\begin{document}
\title{Validating Formal Specifications with LLM-generated Test Cases}
\titlerunning{Validating Formal Specifications with LLM-generated Test Cases}
%
\author{Alcino Cunha\inst{1} \orcidID{0000-0002-2714-8027} \and
Nuno Macedo\inst{1}\orcidID{0000-0002-4817-948X}}
\authorrunning{A. Cunha and N. Macedo}
%
\institute{INESC TEC \& Universidade do Minho \\ 
Braga, Portugal \\ 
\email{\{alcino,nmacedo\}@di.uminho.pt}}
\maketitle              
\begin{abstract}
Validation is a central activity when developing formal specifications. Similarly to coding, a possible validation technique is to define upfront test cases or scenarios that a future specification should satisfy or not. Unfortunately, specifying such test cases is burdensome and error prone, which could cause users to skip this validation task. This paper reports the results of an empirical evaluation of using pre-trained large language models (LLMs) to automate the generation of test cases from natural language requirements. In particular, we focus on test cases for structural requirements of simple domain models formalized in the Alloy specification language. Our evaluation focuses on the state-of-the-art GPT-5 model, but results from other closed- and open-source LLMs are also reported. The results show that, in this context, GPT-5 is already quite effective at generating positive (and negative) test cases that are syntactically correct and that satisfy (or not) the given requirement, and that can detect many wrong specifications written by humans.
\end{abstract}

\section{Introduction}

Formal methods practitioners mainly focus on verification (``did we build the software right?''), and tend to neglect the equally important validation (``did we build the right software?''). Validation aims to ensure that formal specifications are an accurate representation of the system and its requirements. Sitting at the border between the formal and the informal, and requiring the judgment of human stakeholders, validation cannot be fully automated and thus several, often complementary, techniques can be adopted. Simulation is popular in the validation of formal models of behavior, such as state machines, and is available in most model-checkers. Non-essential, even trivial, assertions can be verified to make sure the model is free of specific bugs (``check everything you can''~\cite{lamport2002specifying}). Inspired by test-driven development for coding, \emph{test-driven modeling}~\cite{zugal2012toward} can be adopted, where test cases and scenarios are defined before the formal specifications and later used to validate them. Some advocate that \emph{validation obligations}~\cite{mashkoor2021validation} should be systematically incorporated in the development process, to be discharged as the specification evolves, likewise \emph{proof obligations} for verification.

This paper focuses on the use of test cases for the validation of formal specifications of structural (non-behavioral) requirements of domain models. It focuses on the Alloy~\cite{jackson2012software,jackson2019alloy} specification language, since it is particularly well suited for \emph{i}) the formalization of domain models due to its ``everything is a relation'' design principle; and \emph{ii}) the employment of a test-driven approach. Alloy allows the automatic generation of examples satisfying (or not) a given formal specification, and their graphical depiction with user-customizable themes. This provides good support for validation, as examples can be shown to domain experts to help detect specification errors or clarify ambiguous requirements. 
However, this approach is not suitable for the iterative development of specifications, as it requires constant human intervention. In those contexts, it is preferable to adopt a test-driven modeling approach, where suites of executable (positive and negative) test cases are defined upfront with the help of domain experts, which can then be automatically run to validate specifications as they evolve. Alloy can support this methodology, since models can contain test cases as executable analysis commands. In fact, previous work has leveraged Alloy test cases for various tasks, such as AUnit~\cite{sullivan2018aunit}, a test automation tool that provides coverage criteria, and can also generate tests suites or perform mutation testing~\cite{sullivan2017automated}, or ARepair~\cite{wang2019arepair}, an automatic specification repair technique that uses test cases as oracle.

However, it is difficult to devise a good test suite with sufficiently diverse test cases, and despite the Alloy support for their specification and execution, it is still rather laborious and error-prone to encode them. Existing techniques for automatic test case generation (e.g.,~\cite{sullivan2017automated}), target the debugging phase and already require a (buggy) formal specification. These do not apply to our test-driven approach, where test cases are specified in early development phases from the natural language requirements. 
LLMs have been used successfully to generate unit tests for programs, as shown by various evaluation studies (e.g., \cite{schafer2023empirical,yang2024evaluation}). Those experiments and their results do not directly transfer to our setting, where the input is a natural language requirement (instead of method implementations) and the quality of the generated test suite cannot be measured by code coverage.
In this context, we devised a study to empirically evaluate whether LLMs could also be used to successfully generate test cases to validate formal specifications. Thus, our main contributions are:
\begin{itemize}
    \item The design of the first study to evaluate the effectiveness of LLMs in generating test suites for the validation of formal specifications of structural requirements in domain modeling. It targets Alloy and GPT-5, a state-of-the-art LLM, using a public benchmark for domain models specified in Alloy.
    \item The execution of the study and an evaluation of the results under different aspects, namely: what is the influence of prompt design? what is the effect of non-determinism? how do competing LLMs perform? what characterizes invalid test cases? how diverse are the test suites?
    \item A summary of the major findings and implications for future research.
\end{itemize}
The scripts, raw data, and analysis results of our study are publicly available in the GitHub  repository at \url{https://github.com/haslab/Alloy-LLM-Testing}.

\section{Test-driven modeling with Alloy}
\label{sec:alloy}

Alloy~\cite{jackson2012software,jackson2019alloy} is a formal specification language initially focused on the design and specification of software structures\footnote{Alloy 6 also allows the declarative specification of behavioral requirements using temporal logic~\cite{macedo2016lightweight}. However, our experiment focuses on structural assumptions, so the presentation is restricted to a subset of the language compatible with Alloy 5.}.
In Alloy ``everything is a relation'', meaning all concepts in a domain are modeled with (mathematical) relations, sets of tuples of a given arity. Relations of arity 1 represent sets of elements and relations of higher arity represent associations between those elements. Declared sets are known as \emph{signatures} (keyword \a{sig}). Top-level signatures partition the universe of discourse into disjoint types, but it is also possible to declare arbitrary or disjoint sub-set signatures (keywords \a{in} or \a{extends}, respectively). A signature can also be marked as \a{abstract}, forcing it to only contain elements that are contained in its extensions.
Figure~\ref{fig:alloyexample} presents the Alloy encoding of a domain model for a course management system, one of the examples used in our evaluation. In this model we have four top-level signatures (\a{Person}, \a{Course}, \a{Project}, and \a{Grade}), encoding the key concepts in the domain, and two sub-set signatures of \a{Person}, \a{Professor}s and \a{Student}s. These two signatures are arbitrary sub-sets because each person can simultaneously be a professor and a student (or neither).
The automatic analysis of Alloy models (namely, checking assertions) is done via a bounded model finding procedure, so in analysis commands all top-level signatures have a \emph{scope} that sets the maximum number of elements they can contain. 

\begin{figure}[t]
    \centering
\begin{adjustbox}{scale=0.9}
\begin{alloyfig}
open util/ordering[Grade]
sig Person {
  teaches : set Course,
  enrolled : set Course,
  projects : set Project 
}
sig Professor, Student in Person {}
sig Course { 
  projects : set Project,
  grades : Person -> Grade 
}
sig Project {}
sig Grade {}

fact Enrollment { 
  // Only students can be enrolled in courses
    
}
\end{alloyfig}
\end{adjustbox}
    \caption{Courses domain model in Alloy}
    \label{fig:alloyexample}
\end{figure}

\emph{Fields} are relations of arity higher than 1 declared inside a signature. Our example has five fields: \a{teaches}, \a{enrolled}, and \a{projects} associate each person with, respectively, the courses it teaches, the courses where it is enrolled, and the projects it is working on; (a different) \a{projects} and \a{grades} associate each course with, respectively, the projects proposed by it and the grades obtained by each person in it. Alloy supports overloading (there are two fields named \a{projects}) and fields of arity higher than 2 (\a{grades} is a ternary relation).
Alloy has a module system, and the most frequently used among the pre-defined modules is \a{util/ordering}, a parametrized module that imposes a total order on a given signature. In particular, it declares a binary relation \a{next} that for each element gives the next element in the total order. 
The example uses it on signature \a{Grade}, so that grades can be compared to determine which one is higher. Relation \a{next} is also overloaded, since several total orders can co-exist in a model.

After declaring the relations that capture the entities and associations of a domain model, we proceed to specify additional requirements on those structures. Typically, these are elicited with domain experts using natural language prior to formalization. In Alloy, assumptions about the domain are specified inside \a{fact}s using relational logic, an extension of first-order logic with relational operators (such as dot-join composition or transitive closure), allowing a rather concise specification style. Additionally, \a{pred}icates and \a{fun}ctions can be declared for auxiliary formulas and expressions, respectively. Our example has one (empty) \a{fact} declared, \a{Enrollment}, annotated with a natural language assumption.

\begin{figure}[t]
    \centering
\begin{adjustbox}{scale=0.9}
\begin{alloyfig}
run Positive {
// Only enrolled person is a Student
  some disj P1 : Person, disj C1 : Course, disj G1,G2 : Grade {
    Person = P1
    Professor = none
    Student = P1
    Course = C1
    Project = none
    Grade = G1 + G2
    teaches = none->none
    enrolled = P1->C1
    Person <: projects = none->none
    Course <: projects = none->none
    grades = none->none->none
    Grade <: next = G1->G2 
  }
} for 1 Person, 1 Course, 0 Project, 2 Grade expect 1

run Negative { 
// A professor (not a student) is enrolled in multiple courses
  some disj P1,S1 : Person, disj C1,C2 : Course, disj G1,G2 : Grade {
    Person = P1 + S1
    Professor = P1
    Student = S1
    Course = C1 + C2
    Project = none
    Grade = G1 + G2
    teaches = none->none
    enrolled = P1->C1 + P1->C2 + S1->C1
    Person <: projects = none->none
    Course <: projects = none->none
    grades = C1->S1->G1
    Grade <: next = G1->G2 
  }
} for 2 Person, 2 Course, 0 Project, 2 Grade expect 0
\end{alloyfig}
\end{adjustbox}
    \caption{Example test cases for ``Only students can be enrolled in courses''}
    \label{fig:testcases}
\end{figure}

Besides analysis commands to check assertions, Alloy also supports so-called \a{run} commands, that can be used to search for model instances. This enables test-driven modeling in Alloy, since such commands allow for the specification of test cases to be used for validation. Figure~\ref{fig:testcases} presents two such test cases, one positive that satisfies the assumption, and one negative, that does not satisfy it. It is important to specify both positive and negative test cases: the former allows the detection of over-specifications (if too restrictive, valid instances may be excluded), and the latter under-specifications (if too permissive, invalid instances may be accepted).
Each \a{run} command encodes a specific test case, in the sense that it is a complete valuation of the declared signatures and fields, denoting one concrete instance of the model. They can be specified in relational logic using the well-known \a{some disj} pattern~\cite{practicalalloy}, where an outermost existential quantifier with the \a{disj} keyword denotes the different elements of each signature, followed by a formula that specifies the value of each signature and field using equality constraints. On the right-hand-side of each equality we have an expression that denotes a set of tuples, specified as a union (\a{+}) of singleton tuples, each specified with the cross product operator (\a{->}). An empty signature is equal to \a{none} (a constant denoting an empty set of arity 1), and an empty field equal to an appropriate cross-product of \a{none} (although still denoting an empty set, to avoid an arity type error both sides of the equality must have the same arity). To disambiguate overloaded fields, the domain restriction operator (\a{<:}) is used. A \a{run} command should also specify the scope for each top-level signature after the \a{for} keyword, and the expectation about its satisfiability: for a positive test case we should use \a{expect 1} (it should yield an instance, i.e., be satisfiable); in negative test cases, \a{expect 0} should be used.

Following a test-driven modeling approach, after specifying test cases we proceed to the formalization of the assumption in \a{fact} \a{Enrollment}. Let us suppose a (wrong) attempt at specifying the assumption was the following. The dot-join operator (\a{.}) is used to determine the set of courses where \a{p} is enrolled in.
\begin{alloy}
all p : Person | p in Student implies some p.enrolled    
\end{alloy}
The Alloy Analyzer has an option to run all analysis commands in sequence, reporting a summary of the respective results at the end. Running our two test cases would immediately report that our negative test case did not fail, contrary to expectation. The reason is that this specification still allows professors that are not students to also be enrolled in courses.
If we change the specification to a correct one, for example by swapping both sides of the \a{implies}, both test cases would have the expected outcome. However, the following wrong specification would still pass the two test cases. 
\begin{alloy}
all p: Professor | no p.enrolled    
\end{alloy}
This is the most common wrong specification written by our students for this requirement, because some incorrectly assume that professors cannot be simultaneously students. A test suite with diverse test cases is essential to help the specifier quickly rule out many wrong specifications.
Nonetheless, note that the test-driven modeling approach alone might not be sufficient to ensure proper validation, since it is difficult to devise test cases that catch subtle errors. As such, it should be complemented by other validation tasks, such as checking expected (even apparently trivial) derived assertions.

\section{Study design}
\label{sec:studydesign}

It should be now clear that, although Alloy can support test-driven modeling, it is rather laborious and error-prone to specify a diverse test suite manually. 
With this in mind, and following the recent successes in using LLMs for various software development and formal specification tasks, we devised an empirical evaluation study to assess the effectiveness of a state-of-the-art LLM for generating test suites for validating formal specifications. In particular, we studied version 2025-08-07 of GPT-5, one of the latest and most powerful models offered by OpenAI. 
In fact, the two test cases presented in the previous section were generated by GPT-5 in one of the experiments in our study. This section presents the research questions (RQs) and the design of the study, while the results are presented and discussed in the next section.

\subsection{Research questions}

The proposed empirical evaluation study aims to answer the following RQs:
\begin{description}
    \item[RQ1] How does prompt design influence the effectiveness of test case generation? In particular, what is the effect of including examples in the prompt, comparing zero-shot, one-shot, and few-shot prompts.
    \item[RQ2] What is the effect of non-determinism in test case generation? Even with low temperature settings, LLMs act non-deterministically, giving different responses to the same input, which may affect the outcomes.
    \item[RQ3] How do different LLMs compare in test case generation? Our evaluation focused mainly on GPT-5, but we also compared its effectiveness with other state-of-the-art LLMs, as well as cost-effective and open-source small models.
    \item[RQ4] What are the characteristics of invalid tests cases? Test cases can be invalid for different reasons (from invalid syntax to failure to comply with the requirement), so we studied the characteristics of invalid test cases to determine which factors are currently more challenging for LLMs.
    \item[RQ5] How good are the generated test suites at finding incorrect specifications? The previous RQs focus on the syntactic and semantic validity of generated test suites. This RQ aims to assess the diversity of the test cases through their ability to detect wrong formalizations of the natural language requirements.
\end{description}

\subsection{Benchmark}

For our evaluation we used four small domain models that are part of the Alloy4Fun dataset. Alloy4Fun~\cite{macedo2021experiences} is a web-based version of Alloy that, among other features, enables the creation of small specification challenges whose correctness is checked against a hidden oracle.
Alloy4Fun is used in some universities to help students study Alloy autonomously, including in our own formal methods courses, and over the years we collected thousands of correct and incorrect attempts at specifying several requirements. Most of this data is publicly available in a Zenodo data-set~\cite{alloy4fun_dataset}, and has been regularly used in the evaluation of Alloy-related techniques and tools.

An Alloy4Fun exercise usually consists of a simple model, declaring a few signatures and fields, and a set of specification challenges, each a predicate annotated with a natural language requirement that the student should specify. When specifying a requirement, the user can assume that all previous ones are valid. Table~\ref{tab:becnhmark} describes the four exercises that we used for the evaluation, including the respective links to the Alloy4Fun web-application. Besides the number of requirements in each example, it also gives some statistics about the number of non-equivalent attempts at formalizing each requirement. To determine the latter, we ran a script over the data-set, to group all submissions to each requirement by logic equivalence~\cite{a4f_semantic_dataset}. In the paper GitHub repository we include the results of that analysis, including for each wrong specification of each requirement the syntactic formulation that occurred more frequently. 

\begin{table}[t]
    \centering
    \caption{Overview of the examples used for evaluation}
\label{tab:becnhmark}
    \begin{tabular}{l|c|c|c|c}
        \textbf{Example} &  \textbf{Reqs} & \multicolumn{3}{c}{\textbf{Wrong specs}} \\
         & & \textbf{Min} & \textbf{Max} & \textbf{Mean} \\
        \hline \hline
         \href{http://alloy4fun.inesctec.pt/u6EPDxzJrQSEKRwSf}{Social network} & 8 & 44 & 848 & 272\\
         \href{http://alloy4fun.inesctec.pt/n5rgKi6xhjymixNGy}{Production line} & 10 & 28 & 585 & 147\\
         \href{http://alloy4fun.inesctec.pt/6ymnREeGQGRZFGcyz}{Train station} &  10 & 36 & 203 & 104 \\
         \href{http://alloy4fun.inesctec.pt/Z3SaBk4fvFEicsWwc}{Courses} & 15 & 33 & 312 & 137 \\
         \hline
            & 43 & 28 & 848 & 157
    \end{tabular}
\end{table}

\begin{table}[t]
    \centering
    \caption{Requirements of the courses domain model}
    \label{tab:coursesrequirements}
    \begin{tabular}{l|c}
        \textbf{Requirement} & \textbf{Wrong} \\
        \hline \hline
        Only students can be enrolled in courses & 161 \\
        Only professors can teach courses & 33 \\
        Courses must have teachers & 82 \\
        Projects are proposed by one course & 87 \\
        Only students work on projects and~\ldots & \\
	    \ldots~projects must have someone working on them & 248 \\
        Students only work on projects of courses they are enrolled in & 168 \\
        Students work on at most one project per course & 177 \\
        A professor cannot teach herself & 79 \\
        A professor cannot teach colleagues & 312 \\
        Only students have grades & 80 \\
        Students only have grades in courses they are enrolled & 87 \\
        Students have at most one grade per course & 87 \\
        A student with the highest mark in a course must~\ldots \\
        \ldots~have worked on a project on that course & 291 \\
        A student cannot work with the same student in different projects & 51 \\
        Students working on the same project in a course cannot~\ldots & \\
        \ldots~have marks differing by more than one unit & 109 
    \end{tabular}    
\end{table}

For illustration purposes, consider the courses domain model already presented in the previous section. As we have seen, this example uses a few advanced Alloy features, namely non-binary fields, overloading, and the \a{util/ordering} module. It contains 15 specification challenges, whose natural language description can be seen in Table~\ref{tab:coursesrequirements}. 
The table also shows the number of (semantically distinct) wrong specifications of each requirement, which vary between 33 for requirement ``Only professors can teach courses'', and 312 for requirement ``A professor cannot teach colleagues'', with an average of 137 per requirement. A higher number of wrong specifications can be a sign of both complexity and ambiguity. However, although some requirements may seem ambiguous, we argue that the one that is codified in the hidden oracle is always the most ``correct'' interpretation.
Take, for example, the requirement with the most number of wrong specifications, ``A professor cannot teach colleagues''. At first, many students seem to interpret ``colleagues'' as any other professor. Under the previous requirements (which should be assumed to be true), that would forbid students that are also professors to be enrolled in any course. Since professors that are also students should be able to enroll, the reasonable interpretation of ``colleagues'' in this context is the one specified in the oracle: professors that teach some course together.
The three other domain models are the following:
\begin{itemize}
    \item A photo sharing social network model, where some users are influencers and some photos are ads. This is a simple example in terms of Alloy features (e.g., all fields are binary).
    However, it is the example with the highest average of wrong specifications per requirement. This could be either because it is the first example we give to students in our courses, so it is natural that they make more mistakes, or because some of the requirements are perhaps not standard in real social networks, which might confuse students.
    \item A production line model, that includes both human and robot workers, and that assembles complex, potentially dangerous, components that are made of other components and/or raw materials. This is the most complex example in terms of signature hierarchy, using both signature inclusion and extension, and signature multiplicities. 
    \item A train station railway network model where tracks in the station can be entries, exits, or junctions, and where tracks can have signals attached. This is the simplest example. 
\end{itemize}
The natural language requirements can be consulted by following the links in Table~\ref{tab:becnhmark}, and are also available in the GitHub repository, along with the oracles.

\subsection{Prompt design}

Most LLMs nowadays distinguish between \emph{system} and \emph{user} prompts: the former clarify the instructions an LLM should follow in the subsequent interactions (for example, defining roles or formatting guidelines for the responses), while the latter contain task requests or questions to be answered according to the system instructions. We used both prompts in our evaluation. The user prompt is simple, asking for test cases for a given requirement and model. Let $R_i$ be the natural language formulation of the $i$-th requirement of a domain model, and $N$ be the number of positive and negative test cases to be generated. The user prompt starts as follows, after which the domain model in Alloy is provided.
\lstset{
  basicstyle=\small\ttfamily,
  mathescape
}
\begin{lstlisting}
Generate $N$ positive and $N$ negative instances for the  requirement "$R_i$" 
for the following model. All instances must also satisfy the 
requirements "$R_0$", $\ldots$, and "$R_{i-1}$".
\end{lstlisting}
Minor adjustments are made for $R_0$ (removing the second sentence) and $R_1$ (adjusting the second sentence to a single requirement).

Since we want to measure the effect of including examples, we created three versions of the system prompt: \emph{zero-shot}, \emph{one-shot}, and \emph{few-shot}. The three versions were iterated several times in the GPT-5 console until we reached a satisfactory result. These were then fixed and used for the whole evaluation with all LLMs, which was performed via remote calls to the respective APIs. All three system prompts start with a brief presentation of Alloy (a couple of sentences) and end with a definition of the LLM role and an itemized list of precise instructions.
The main differences are the following:
\begin{description}
    \item[Zero-shot] Besides the mentioned Alloy introduction and final list of instructions, this version has no Alloy code, including only a paragraph in natural language explaining that test cases can be encoded using \a{run} commands with existential quantifiers and equality constraints. 
    \item[One-shot] Besides the content of the \emph{zero-shot} prompt, this version includes an example of an Alloy domain model (different from the ones used in the evaluation) that uses all relevant features of the language, a paragraph explaining this model and the respective features, and a single example of a test case specified in Alloy using the \a{some disj} idiom.
    \item[Few-shot] This version introduces Alloy features incrementally and provides examples of test cases along the way. It starts with a simple version of an Alloy domain model (just with two signatures and a binary field) along with a positive and a negative test case encoded with the \a{some disj} idiom, and a test case explaining how to specify that a field is empty. Then, signature extension and inclusion are introduced with a variant of the initial model, along with a test case example. Finally, ternary relations and the \a{util/ordering} module are introduced, along with a final test case example. In total, this prompt has three variants of a domain model and five examples of test cases.
\end{description}
The three variants of the system prompt can be found in the GitHub repository. 

\subsection{Metrics}

For the first three RQs we evaluated the effectiveness in terms of syntactic and semantic correctness of the generated test suite. In particular, our result tables will present the following metrics:
\begin{description}
    \item[Tests] The number of test cases to be generated. We ask the LLM to generate $N$ positive and $N$ negative test cases for each requirement. As such, the full test suite should have $2 \times N \times 43$ test cases, since we have 43 requirements in total.
    \item[Syntax] The number of test cases that are syntactically correct. To measure this, we attempt to parse each \a{run} command in isolation together with the respective Alloy model. 
    \item[Consistent] The number of syntactically correct test cases that produce an instance. A syntactically correct \a{run} command can be inconsistent for various reasons (e.g., wrongly specified scopes). To measure this, we execute each command in isolation together with the respective Alloy model and check if the result is satisfiable.
    \item[Previous] The number of executable test cases that satisfy all previous requirements. To measure this we add to the model facts with the oracles of all previous requirements and check if each \a{run} command is still satisfiable.
    \item[Valid] The number of test cases satisfying previous requirements that agree with the specified oracle. To measure this we add the oracle specification as a \a{fact} to the model, and check if the \a{run} command has the  expected outcome (satisfiable/unsatisfiable for positive/negative tests). We also report this metric as a percentage of the total number of generated tests (\textbf{\%}).
    \item[Cost] The actual cost (at the date of the study) of using the LLM to generate the test suite. This is an indicator of the number of tokens used in total, including input, output, and reasoning tokens. 
\end{description}

For RQ5 we evaluated the effectiveness in terms of the ability of the test suite to actually catch wrong specifications. To make fair comparisons, we focus only on the requirements for which the LLM generates a full test suite, with $2 \times N$ valid test cases. In the results for this RQ we present the following metrics:
\begin{description}
    \item[Complete] The number of requirements for which the LLM generates a complete valid test suite (out of the 43 possible).
    \item[Wrong] The total number of incorrect specifications, considering only  the requirements with a complete test suite.
    \item[Missed] The total number of incorrect specifications that the test suite can detect. To measure this, we run all $2 \times N$ test cases of each requirement against each wrong specification and checked if the result was the expected. 
    \item[Mean \%] The average of missed wrong specifications per requirement. Note that, since results are presented grouped by run, this is not the same as the ratio between \textbf{Missed} and \textbf{Wrong}.
\end{description}

\section{Results}
\label{sec:results}

This section presents a summary of the study results. The fully detailed results per example and per requirement can be found in the GitHub repository.

\subsection{RQ1: influence of prompt design}

To assess the influence of prompt design we asked GPT-5 to generate test suites using the three different prompt versions described above with $N=3$. The results are presented in Table~\ref{tab:promptresults}. As can be seen, the \emph{few-shot} prompt clearly outperforms the other two versions, being able to generate 247 valid tests out of the 258 requested (a success rate of 96\%). Of the 11 test cases that were not valid, 3 failed the syntax check, 3 failed to satisfy the previous requirements, and 5 did not give the expected result according to the oracle. The 11 invalid test cases were distributed among 7 requirements, but all still had at least 3 valid test cases.
The success rate drops to 79\% for the \emph{one-shot} prompt and 46\% for the \emph{zero-shot}. Most of this decrease is due to syntactic problems: among the test cases that are executable, 91\% and 86\% satisfy the previous requirements and the oracle with \emph{one-shot} and \emph{zero-shot} prompts, respectively. A possible explanation is the lack of Alloy test case examples in the training data, since most Alloy publications and educational material do not adopt a test-driven methodology. 
Interestingly, although the input size is much smaller for prompts with fewer examples, the total cost to generate the test suite is higher. The pricing structure for most commercial LLMs imposes a higher price on output tokens when compared to input tokens (sometimes $10\times$ more), and reasoning tokens are charged as output tokens. Since the output is of comparable size, having less information and examples in the system prompt led to the need for more reasoning tokens, which alone justifies the higher cost.

\begin{table}[t]
    \centering
    \caption{Influence of prompt design}
    \label{tab:promptresults}%
    \begin{tabular}{l|c|c|c|c|c|c|c}
        \textbf{Prompt} &  \textbf{Tests} & \textbf{Syntax} & \textbf{Consistent} & \textbf{Previous} & \textbf{Valid} & \textbf{\%} & \textbf{Cost} \\
        \hline \hline
         \textbf{Few} & 258 & 255 & 255 & 252 & 247 & 96\% & \$3.56\\
         \hline 
         \textbf{One} & 258 & 226 & 208 & 206 & 205 & 79\% & \$3.69 \\
         \hline 
         \textbf{Zero} & 258 & 137 & 120 & 119 & 118 & 46\% & \$4.20\\
    \end{tabular}
\end{table}
\begin{table}[t]
    \centering%
    \caption{Effect of non-determinism}%
    \label{tab:nondeterminismresults}
    \begin{tabular}{l|c|c|c|c|c|c|c}
        \textbf{Run} &  \textbf{Tests} & \textbf{Syntax} & \textbf{Consistent} & \textbf{Previous} & \textbf{Valid} & \textbf{\%} & \textbf{Cost} \\
        \hline \hline
         \textbf{1st} & 258 & 255 & 255 & 252 & 247 & 96\% & \$3.56\\
         \hline 
         \textbf{2nd} & 258 & 256 & 256 & 256 & 251 & 97\% & \$3.50\\
         \hline 
         \textbf{3rd} & 258 & 252 & 252 & 250 & 246 & 95\% & \$3.61\\
    \end{tabular}
\end{table}

\begin{center}
\fbox{\begin{minipage}{0.95\textwidth}
\textbf{Finding 1}: GPT-5 with a \emph{few-shot} system prompt is highly effective at generating Alloy test suites for structural requirements expressed in natural language. It is also more cost-effective due to less reasoning effort.
\end{minipage}}
\end{center}

\subsection{RQ2: effect of non-determinism}

LLMs are inherently non-deterministic, and can give different results for identical inputs. Some allow the user to set the so-called \emph{temperature}, a parameter that affects the randomness of results, by adjusting the probability distribution of the next output token. However, even with temperature set to 0 (the most deterministic setting) determinism is not guaranteed due to hardware factors like GPU concurrency. GPT-5 in particular does not allow the user to set this parameter, so it is even more relevant to study the impact of non-determinism on the results. To assess that effect, we run GPT-5 three times with the same task ($N=3$), and using the \emph{few-shot} prompt that proved to be more effective. The results are presented in Table~\ref{tab:nondeterminismresults} (the \textbf{1st run} is the \textbf{Few-shot} run presented in Table~\ref{tab:promptresults}). As can be seen, the overall results do not vary significantly, with a success rate always close to 96\%. For 33 of the 43 requirements we always got the same number of valid, albeit possibly different, test cases.

\begin{center}
\fbox{\begin{minipage}{0.95\textwidth}
\textbf{Finding 2}: The overall effectiveness of GPT-5 when using a \emph{few-shot} prompt does not seem to be significantly affected by non-determinism.
\end{minipage}}
\end{center}

\subsection{RQ3: LLM comparison}

To answer this RQ, we compared the effectiveness of GPT-5, version 2025-08-07, with two other state-of-the-art models, Gemini 2.5 Pro from Google DeepMind and Claude Opus 4.1, version 2025-08-05, from Anthropic. We also compared with two less advanced models, namely GPT-5 Mini, version 2025-08-07, a cost-efficient and faster version of GPT-5, also from OpenAI, and the open-source smaller language model Llama 3.1 8B from Meta. Table~\ref{tab:llmresults} presents the results, again with $N=3$ and the \emph{few-shot} prompt. The results in \textbf{GPT-5} are the ones corresponding to the \textbf{Few-shot} run presented in Table~\ref{tab:promptresults}. The results from Llama are omitted because this model performed quite badly, and was only able to generate a handful of syntactically correct test cases. Further research is needed to determine if small language models in general are unable to properly cope with this task, or if this result is specific to Llama. The two competing state-of-the-art models performed worse than GPT-5, although still with a success rate above 75\%. Gemini 2.5 Pro struggled more with getting test cases that are executable (for example, forgetting to set the scope for a signature), while Claude Opus 4.1 always succeeded in that aspect but struggled to generate test cases that satisfied previous requirements. The main problem for the cost-efficient GPT-5 Mini was getting the syntax right, a pattern similar to the one obtained with GPT-5 for simpler the prompts.

\begin{table}[t]
    \centering
    \caption{LLM comparison}
    \label{tab:llmresults}
    \begin{tabular}{l|c|c|c|c|c|c|c}
        \textbf{Model} &  \textbf{Tests} & \textbf{Syntax} & \textbf{Consistent} & \textbf{Previous} & \textbf{Valid} & \textbf{\%} & \textbf{Cost} \\
        \hline \hline
         \textbf{GPT-5} & 258 & 255 & 255 & 252 & 247 & 96\% & \$3.56\\
         \hline 
         \textbf{Gemini} & 258 & 248 & 220 & 212 & 210 & 81\% & \$2.78\\
         \hline 
         \textbf{Claude} & 258 & 258 & 258 & 202 & 197 & 76\% & \$5.55\\
         \hline 
         \textbf{Mini} & 258 & 188 & 183 & 179 & 174 & 67\% & \$0.53\\
    \end{tabular}
\end{table}
\begin{table}[t]
    \centering
    \caption{Wrong specification detection}
    \label{tab:errordetectionresults}
    \begin{tabular}{l|c|c|c|c}
        \textbf{$N$} & \textbf{Complete} & \textbf{Wrong} & \textbf{Missed} & \textbf{Mean \%} \\
        \hline \hline
        \textbf{1} & 41 & 5587 & 2204 & 38.10\%\\
        \hline 
        \textbf{2} & 34 & 4623 & 814 & 17.02\%\\
        \hline 
        \textbf{3} & 36 & 4395 & 525 & 9.90\%\\
        \hline 
        \textbf{4} & 35	& 5129 & 446 & 7.45\%\\
        \hline 
        \textbf{5} & 35	& 4816 & 348 & 6.43\%\\
    \end{tabular}
\end{table}

\begin{center}
\fbox{\begin{minipage}{0.95\textwidth}
\textbf{Finding 3}: Different LLMs struggle with different aspects of test case generation, some struggle more with syntax while others with semantics.
\end{minipage}}
\end{center}

\subsection{RQ4: characterization of invalid test cases}

To answer this RQ we manually analyzed all invalid test cases in the 3 runs of GPT-5 reported in Table~\ref{tab:nondeterminismresults}. In total we had 30 invalid test cases, distributed as follows: 11 were syntactically incorrect, 5 failed to satisfy previous requirements, and 14 did not agree with the oracle. 

All the 11 syntactically incorrect test cases were due to the same problem: to specify that a binary relation \a{R} is empty, formula \a{R = none} was used instead of \a{R = none->none}. This is a reasonable mistake to be made (often made by our students also), since the latter syntax is an Alloy specific notation needed to make the expression type-check. We actually noticed this problem in the initial experiments with the prompt design, and explored the viability of the alternative (uniform) syntax where to state that \a{R} is empty one just specifies \a{no R}. However, the LLM also struggled with this formulation, since it deviated from the standard equality constraint, sometimes generating the also syntactically incorrect \a{R = no R}. These are simple syntactic errors, that could easily be addressed with postprocessing the result.

\vspace{0.2cm}
\begin{center}
\fbox{\begin{minipage}{0.95\textwidth}
\textbf{Finding 4}: The syntactic errors made by GPT-5 are due to a peculiar feature of Alloy syntax and easily fixable by some post-processing.
\end{minipage}}
\end{center}
\vspace{0.2cm}

Of the 5 test cases that failed previous requirements, one was because the test case was not fully specified and admitted multiple instances, some of them valid and some invalid. No pattern was detected in the remaining 4, all of them indeed failed in one of the previous requirements that were stated in the task prompt. 
Strangely, in 3 cases where the generated test cases for $R_i$ failed to satisfy a previous requirement $R_j$ for $j < i$, the LLM had been able to generate valid test cases directly for $R_j$.

Of the 14 test cases that did not agree with the oracle, 13 were negative test cases, which seems to suggest that the LLM struggles more with these. The only positive test case that was invalid was for the requirement ``Components cannot be their own parts'' in the production line example. The invalid test case had a component \a{A} that was a part of component \a{B} which in turn was part of \a{A}. Essentially, the requirement was interpreted literally, but obviously it is a wrong interpretation, as components cannot physically contain themselves direct or indirectly. 
The 13 invalid negative test cases were distributed among only 3 requirements: 1 for the requirement ``Only students work on projects and projects must have someone working on them'' in the courses example, because a student that is also a professor was wrongly assumed to not be able to work on a project; 3 for the requirement ``Users can see ads posted by everyone, but only see non ads posted by followed users'' in the social network example, due to a misinterpretation of the ``can'' as a ``must'' (all test cases had a user that did not see an ad, which was considered to be wrong); and 9 for the requirement ``A professor cannot teach colleagues'' because, likewise many of our students, it was incorrectly assumed that ``colleagues'' meant any other professor. Note that all 3 runs of GPT-5 made the last two mistakes, which is another indicator that the results are not significantly affected by non-determinism.

\begin{center}
\fbox{\begin{minipage}{0.95\textwidth}
\textbf{Finding 5}: GPT-5 struggles more with negative test cases then with positive ones, but only on specific requirements that our anecdotal evidence suggests to be also ambiguous to humans.
\end{minipage}}
\end{center}

\subsection{RQ5: effectiveness at detecting incorrect specifications}

For this RQ we asked GPT-5, using the \emph{few-shot} system prompt, to generate test suites of increasing size, from $N=1$ to $N=5$. For the requirements where a complete valid test suite was obtained we computed the ratio of undetected wrong specifications in our benchmark. Table~\ref{tab:errordetectionresults} presents the results, reporting for each test suite size the average of those ratios. As can be seen the ratio of missed wrong specifications decreases as the test suite size increases, with significant drops when $N$ is increased from 1 to 2 and from 2 to 3. This is a clear indicator that GPT-5 is generating different scenarios in the test cases of each test suite.
In the biggest test suite, in average only 6.43\% of the wrong specifications went undetected, but only in 5 requirements were all of them detected. To put these results in context, we know (by using a brute-force exploration) that for most requirements it is possible to have a small test suite that catches all wrong specifications (in most cases it suffices to have $N=2$ with a scope of 3 for top-level signatures). However, devising such a minimal test suite without knowing \emph{a priori} the wrong specifications would be extremely difficult or even impossible for a human expert, so it is unrealistic to expect the LLM to do it.

\begin{center}
\fbox{\begin{minipage}{0.95\textwidth}
\textbf{Finding 6}: GPT-5 generates diverse test cases for most requirements.
\end{minipage}}
\end{center}%

\subsection{Threats to validity}
The threats to \emph{internal validity} mainly lie in our experimental scripts. To minimize this threat both authors reviewed the code, and we also released all scripts and data in the GitHub repository for replication. The threats to \emph{external validity} mainly lie in the used benchmark and LLMs. Unfortunately, we are not aware of any standard benchmark in this context. We acknowledge that the Alloy4Fun dataset mainly consists of educational examples, with domain models that are simpler than realistic ones. However, some of the requirements are already quite complex, and from our own consultancy experience, realistic cases have much more requirements, but they are not necessarily more complex, so we have some confidence that the results will generalize to that setting. As for the LLMs, we evaluated three of the most prominent commercial offerings so we are confident that the results are representative of the state of the art. For cost-efficient, open-source, and small language models, we only evaluated two offerings so we cannot generalize the results. We also acknowledge that our study focuses on Alloy (for the provided domain model and the test case specifications) and structural requirements, so we cannot generalize to other formal specification languages or other kinds of requirements.

The threats to \emph{construct validity} lie mainly in the prompt design, non-determinism, and data leakage. When designing the study we iterated our prompt texts multiple times with GPT-5 until reasonable results were obtained, but we acknowledge that these may not be ideal for all configurations and other LLMs. 
Concerning non-determinism, ideally all experiments would have been run multiple times, but that would significantly raise the evaluation costs. We opted to study the effect of non-determinism effect in a dedicated RQ only with GPT-5 and a specific test suite size and considering only 3 different runs. This experiment showed that GPT-5 is rather consistent with the results, but we acknowledge that non-determinism is still a potential threat, in particular for the results of other models. As for data leakage, the Alloy4Fun dataset is publicly available for some time and may have been used in the training data of the LLMs. However, the dataset does not include test cases and we are also not aware of any other public dataset with test case examples for the Alloy4Fun domain model examples, so we are confident that data leakage is not a relevant threat.

\section{Related work}
\label{sec:relatedwork}
Alloy's support for test-driven methodologies has led to the development of techniques and tools for unit testing, such as the already mentioned AUnit framework~\cite{sullivan2018aunit}. AUnit automates the generation of unit tests~\cite{sullivan2017automated}---specified using the same \a{some disj} pattern used in this study---but it does so by inspecting a formal specification and covering all its sub-expressions. In validation (our target) it is unknown whether the specification conforms to the requirements, so the user would have to manually classify all test cases. We propose instead to generate test cases from the natural language requirements before formalization starts. For the linked list example from~\cite{sullivan2017automated} our technique easily generates a small test suite to validate the requirement ``The list is acyclic'' that detects the incorrect specification illustrated in the paper. 

We are not aware of any research that uses LLMs to generate test suites for Alloy, but they have been used for other Alloy-related tasks. Hong et al.~\cite{hong2025effectiveness} explore using LLMs to generate Alloy specifications from simple requirements expressed in natural language, and conclude that they can be quite effective. Alhanahnah et al.~\cite{alhanahnah2025empirical} explore the effectiveness of LLMs at repairing buggy Alloy specifications, and conclude that a setup that pairs a repair agent with an instructor agent (that uses feedback from the Alloy Analyzer to generate repair prompts), can surpass state-of-the-art symbolic repair techniques. Similar autoformalization techniques have been developed for other formal specification languages and logics. For example, Zhao et al.~\cite{zhao2024nl2ctl} explored the viability of using LLMs to generate CTL specifications from natural language, and Zuo et al.~\cite{zuo2025pat} the generation of behavioral models specified in the PAT formal method~\cite{sun2009pat}.

The work by Pan et al.~\cite{pan2025llm} on the generation of model instances using LLMs shares some similarity to ours, since it focuses on structural aspects. Given an XML description of a UML class diagram and a natural language description of a desired instance, LLMs are used to generate the corresponding XMI instance model. A key difference is that we do not expect natural language definitions of the test cases, but only of a high-level requirement. Moreover, their evaluation focuses only on syntactic correctness. To make the technique feasible with open-source, small LLMs, they use an abstract intermediate representation that is converted to XMI in post-processing. Our approach could employ a similar technique to reduce the rate of syntactic errors in the less advanced LLMs. 

In the testing community there is some work on using LLMs to generate test cases from natural language requirements (rather than from code), that relate to our work (see, e.g.,~\cite{10.1145/3771727} for a comprehensive survey). For example, Korraprolu et al.~\cite{korraprolu2025test} present an empirical study to compare the effectiveness of LLMs at generating test cases for Simulink from natural language requirements. Effectiveness was measured as coverage of the (executable) models implemented afterwards (the correctness of the test cases was not assessed). Test cases are simple input-output pairs of numerical or/and discrete values and generated in an tabular form (later translated manually to executable tests). A similar study was done by Xue et al.~\cite{xue2024llm4fin} for the concrete domain of FinTech software~\cite{xue2024llm4fin}.
Our tests can also be seen as input-output pairs with a Boolean output but much more complex input (a graph of connected elements from the domain of discourse). Our tests are also directly executable without manual intervention and our evaluation focused on the correctness of the tests.

Several works attempted to generate high-level test cases from natural language requirements, in the sense that tests are themselves written in natural language, and thus potentially more accessible for LLMs. 
For example, Masuda et al.~\cite{masuda2025generating} use LLMs to extract not only high-level tests but also test design techniques from documents describing all application requirements. 
Arora et al.~\cite{arora2024generating} use LLMs to generate test scenarios for an Austrian Post application for managing delivery and postage processes. There is also some work (e.g.,~\cite{chang2024llmscenario}) on generating driving scenarios for autonomous driving applications using LLMs.

\section{Conclusion and future work}
\label{sec:conclusion}

Our experimental evaluation indicates that GPT-5 with a \emph{few-shot} prompt is highly effective and consistent at generating test suites from natural language requirements, that can be used for the validation of structural requirements of domain models. Generated test cases are almost always syntactically correct, executable, and comply with the stated requirements. The generated test suite for each requirement is also diverse and is very effective at detecting incorrect specifications made by humans. Competing state-of-the-art LLMs are also effective, but less so, when compared with GPT-5. In the future we plan to experiment with alternative more targeted prompts (for example, task specific system prompts adapted for each domain model) and add a post-processing syntax repair procedure, that combined could increase the effectiveness of the technique with open source small language models, making the technique more accessible, namely in educational contexts. We also plan to evaluate the effectiveness of this approach for generating test suites for validating behavioral requirements, namely requirements to be specified with temporal logic. That would make the technique usable in a wider range of formal methods.

\section*{Acknowledgments}

This work is funded by national funds through FCT – Fundação para a Ciência e a Tecnologia, I.P., under the support UID/50014/2025 (\url{https://doi.org/10.54499/UID/50014/2025}).

\bibliographystyle{splncs04}
\bibliography{bibliography}
\end{document}